\documentclass[11pt]{article}
\usepackage[]{amsmath,amssymb}
\usepackage{graphics,epsfig}

\begin{document}
\date{}
\title{Entanglement and alpha entropies for a massive Dirac field in two dimensions}
\author{H. Casini\footnote{e-mail: casini@cab.cnea.gov.ar}, 
C. D. Fosco\footnote{e-mail: fosco@cab.cnea.gov.ar}
 and 
M. Huerta\footnote{e-mail: huerta@cabtep2.cnea.gov.ar} \\
{\sl Centro At\'omico Bariloche, 
8400-S.C. de Bariloche, R\'{\i}o Negro, Argentina}}
\maketitle

\begin{abstract}
  We present some exact results about universal quantities derived
  from the local density matrix $\rho$, for a free massive Dirac field in
  two dimensions. We first find $ {\rm tr} \rho^n$ in a novel fashion,
  which involves the correlators of suitable operators in the
  sine-Gordon model. These, in turn, can be written exactly in terms
  of the solutions of non-linear differential equations of the
  Painlev\'e $V$ type. Equipped with the previous results, we find the
  leading terms for the entanglement entropy, both for short and long
  distances, and showing that in the intermediate regime it can be
  expanded in a series of multiple integrals. The previous results
  have been checked by direct numerical calculations on the lattice,
  finding perfect agreement.  Finally, we comment on a possible
  generalization of the entanglement entropy c-theorem to the
  alpha-entropies.
\end{abstract}

\section{Introduction}

The trace of the vacuum state projector over the degrees of freedom
corresponding to a spatial region, results in a mixed density matrix
with a non-vanishing `geometric' entropy. This kind of construction was
proposed by some
authors~\cite{'tHooft:1984re,Bombelli:1986rw,Srednicki:1993im,Callan:1994py}
in an attempt to explain the black hole entropy as some sort of
entanglement entropy of the corresponding vacuum state. However, since
the role of gravity cannot be ruled out, this identification is still
a conjecture.
  
In recent years, there has been a renewed interest on the properties
of local reduced density matrices, specially for low dimensional
systems (see for example \cite{
  Calabrese:2004eu,Casini:2004bw,Vidal:2002rm,Latorre:2003kg,
  Latorre:2004pk,Orus:2005jq,vedral,Plenio:2004he}, and references
therein).  This was partially motivated by developments in quantum
information theory and also on the density matrix renormalization
group method in two dimensions~\cite{density}.  These investigations
have made it manifest that the entanglement entropy, as well as other
measures of information for the reduced density matrices of the vacuum
state are interesting quantities by their onw right. Besides, they can
yield a different view on certain aspects of quantum field theory,
since the objects one has to calculate are quite different to the
standard ones.  That means, for example, that there is a different
structure of divergences, and a nice interplay between geometry and UV
behaviour.

In this paper, we study measures of information for the local density
matrix $\rho_A$ for a Dirac fermion in two dimensions.  This is achieved
by tracing the vacuum state over the degrees of freedom outside the
set $A$.  Specifically, we shall consider the $\alpha$-entropies
\begin{equation}
S_{\alpha }(A)=\frac{1}{1-\alpha} \textrm{log} \, \textrm{tr}(\rho_A^{\alpha })\,,
\label{unos}
\end{equation}
 and the entanglement entropy 
\begin{equation} 
S(A)=-\textrm{tr}(\rho_A \log \rho_A) = \lim_{\alpha \to 1} S_{\alpha }(A).
\label{entro}
\end{equation}

These two functions of $\rho_A$ are perfectly well defined on a lattice,
but their continuum limit are plagued by UV divergences.  In two
dimensions, however, the form of those divergences is particularly
simple, being just an additive constant proportional to the logarithm
of the cutoff and to the number of boundary points in $A$. Indeed,
since the divergences have their origin in the UV fixed-point (where
masses can be neglected), this follows from the result for the
conformal case~\cite{Calabrese:2004eu, Holzhey:1994we}.  Remarkably, for
the entanglement entropy, it also follows from the strong
subadditive property of the entropy, when the spatial symmetries are
taken into account~\cite{Casini:2004bw}.

Several universal quantities can be easily obtained from $S(A)$ and
$S_\alpha(A)$.  In particular, when $A$ is a single interval of length $r$,
we define the dimensionless functions
 \begin{equation}\label{tress}
c_{\alpha}(r)\;\equiv\;r \frac{dS_{\alpha}(r)}{dr}\;\;,\;\;\; c(r)\;\equiv\;c_1(r)
\;.
\end{equation}
The function $c(r)$ is always positive and decreasing, and we shall
call it `entropic $c$-function', since it plays the role of
Zamolodchikov's $c$-function in the entanglement entropy $c$-theorem
\cite{Casini:2004bw}.  All the universal information which can be
obtained from $S_{\alpha }(r)$ and $S(r)$ is encoded in $c_{\alpha}(r)$ and
$c(r)$.

For more general sets, formed by several disjoint intervals, universal
quantities can be constructed through the mutual information function
$I$, which for two non intersecting sets $A$, $B$ is given by
\begin{equation}
I(A,B)=S(A)+S(B)-S(A\cup B)\,. \label{mut} 
\end{equation}
Remarkably, $I$ remains finite in higher dimensions\footnote{Formulae
  (\ref{tress}) make sense only when a translation invariant cutoff is
  used. This apparent limitation could be overcome, in two dimensions,
  by defining $c(r)$ through the function {\mbox{$F(A,B)=S(A)+S(B)-S(A\cap B)-S(A\cup B)$}} used
  in~\cite{Casini:2004bw} for overlapping sets. However, $F(A,B)$ is not finite, in general, for 
dimensions greater than 2.}.

The traces of powers of $\rho$ involved in the $\alpha$-entropies, with $\alpha=n\in
{\mathbb Z}$, can be represented by a functional integral with the
fields defined on an $n$-sheeted surface with conical singularities
located at the boundary points of the set $A$.  This integral is quite
difficult to deal with, except for the conformal case, where that
surface can be conveniently transformed.  

We shall use here a novel approach (see section~\ref{sec:method}), whereby the problem in the
n-covered plane is mapped to an equivalent one in which an external
gauge field couples to an $n$-component fermion field (defined on the
plane) and its role is to impose the correct boundary conditions
through its vortex-like singularities.  Then the resulting theory is
bosonized, to express ${\rm tr} \rho^n$ as a sum of correlators of local
operators. 

In the massless case, discussed in section~\ref{sec:massless}, the
bosonized theory becomes a massless scalar field, and thus we could
obtain the entropies explicitly, with results in agreement with the
ones of~\cite{Calabrese:2004eu}. As a by-product we derived a quite
convenient expression for $I(A,B)$.

For the massive case (section~\ref{sec:massive}), the dual theory is
instead a sine-Gordon model at the free fermion point.  In this case,
we still could evaluate ${\rm tr} \rho^n$ exactly for a single interval
of size $r$, by expressing it as a sum of $n$ correlators for
exponential operators.  A method to find that type of correlators was
(fortunately) already available in the literature~\cite{Bernard:1994re}, and it could be
adapted to our case after some minor modifications. The outcome is an
exact expression for ${\rm tr} \rho^n$ in terms of the solutions of
second order non linear differential equations.  Also for the massive
fermion case, we obtained expansions for small and large values of $m
r$ ($m$ is the fermion mass), and by analytical continuation in $n$ we
 calculated the corresponding expansions for the entropy.

This article concludes with a summary of our results (section~\ref{sec:summary})
and a discussion on a possible generalization of the $c$-theorem to the
$\alpha$-entropies. This conjecture naturally suggests itself by the
results of the explicit calculations.

\section{Reduced density matrix for a Dirac fermion}\label{sec:method}
We consider the reduced density matrix for a free Dirac fermion, which
is defined by tracing the vacuum state over the degrees of freedom
lying outside a given set $A$. We specify $A$ as a collection of
disjoint intervals $ (u_{i},v_{i})$, $i=1,...,p$ (see figure $1$).

Following~\cite{Holzhey:1994we}, the density matrix $\rho (\Psi
_{\textrm{in}},\Psi _{\textrm{out}})$ can be written as a functional
integral on the Euclidean plane with boundary conditions $\Psi =\Psi
_{\textrm{in}}$, and $\Psi =\Psi _{\textrm{out}}$ along each side of the
cuts $(u_{i},v_{i})$ 
\begin{equation}
\rho (\Psi _{\textrm{in}},\Psi _{\textrm{out}})=\frac{1}{Z[1]}\int D\Psi e^{-S[\Psi ]}\,,
\end{equation}
where $Z[1]$ is a normalization factor, introduced in order to have $\textrm{tr}\rho =1$.

\begin{figure} [tbp]
\centering
\leavevmode
\epsfxsize=10cm
\bigskip
\epsfbox{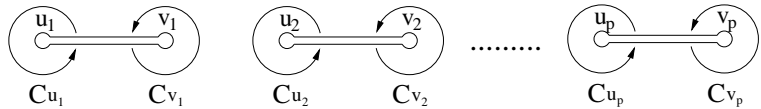}
\caption{The plane with cuts along the intervals $(u_i,v_i)$, $i=1$,...,\,$p$. The circuits 
$C_{u_i}$ and $C_{v_i}$ are used in the text to discuss the boundary conditions.}
\end{figure}

In order to obtain ${\rm tr}\rho^{n}$, we consider $n$ copies of the cut
plane, sewing together the cut $(u_{i},v_{i})_{\textrm{out}}^{k}$ with
the cut $(u_{i},v_{i})_{\textrm{in}}^{k+1}$, for all $i=1,...,p$, and
the copies $ k=1,...,n$, where the copy $n+1$ coincides with the first
one. The trace of $\rho^{n}$ is then given by the functional integral
$Z[n]$ for the field in this manifold,
\begin{equation}
\textrm{tr}\rho^{n}=\frac{Z[n]}{Z[1]^{n}}\,. 
\label{dd}
\end{equation}

For a fermion field, we have to remember that the trace requires to
introduce a minus sign in the path integral boundary condition 
between the fields along the first and the last cut \cite{Larsen:1994yt}, as it happens for the fermion thermal partition
function in the Matsubara formalism~\cite{thermalft}. Besides, in the present case, for each copy there is an additional factor $-1$. This is due to the existence of a non trivial Lorentz rotation around the points $u_{i}$ and $v_{i}$ which is present in the Euclidean Hamiltonian when expressing $\textrm{tr}\rho^n$ as a path integral \cite{kabat}. From these considerations, we finally get a total factor $(-1) ^{(n+1)}$ connecting the fields along the first cut $(u_{i},v_{i})_{\textrm{in}}^{1}$ and the last one $(u_{i},v_{i})_{\textrm{out}}^{n}$ in (\ref{dd}).

Rather than dealing with field defined on a non trivial manifold, we
find it more convenient to work on a single plane, although with an
$n$-component field
\begin{equation}
\vec{\Psi}=\left(\begin{array}{c} 
\Psi _{1}(x) \\ 
\vdots \\  
\Psi _{n}(x) \end{array}
\right) \,,
\label{}
\end{equation}
where $\Psi _{l}(x)$ is the field on the $l^{\textrm{th}}$ copy. Of
course, the singularities at the boundaries are still there, and we
shall show a simple way of taking them into account now.

Note that the space is simply connected but the vector $\vec{\Psi}$
is not singled valued.  In fact, turning around any of the $C_{u_{i}}$
circuits (see figure 1) it is multiplied by a matrix $T$, and after
turning around the $C_{v_{i}}$ circuit it gets multiplied by the
inverse matrix $T^{-1}$. Here,
\begin{equation}
\begin{array}{c}
T=\left(
\begin{array}{lllll}
0 & 1 &  &  &  \\
& 0 & 1 &  &  \\
&  & . & . &  \\
&  &  & 0 & 1 \\
(-1)^{(n+1)} &  &  &  & 0
\end{array}
\right)
\end{array}\,, \label{}
\end{equation}
which has eigenvalues $e^{i\frac{k}{n}2\pi }$, with
$k=-\frac{(n-1)}{2}$, $- \frac{(n-1)}{2}+1$,...,$\frac{(n-1)}{2}$.  

Then, changing basis by a unitary transformation in the replica space,
we can diagonalize $T$, and the problem is reduced to $n$ decoupled
fields $\Phi^{k}$ living on a single plane. These fields are multivalued,
since when encircling $C_{u_{i}}$ or $C_{v_{i}}$ they are multiplied
by $e^{i\frac{k}{n}2\pi }$ or $e^{-i\frac{k}{n}2\pi }$, respectively.

That multivaluedness can now be easily disposed of, at the expense of
coupling {\em singled-valued\/} fields $\Phi^{k}$ to an external gauge
field which is a pure gauge everywhere, except at the points $u_{i}$ and
$v_{i}$ where it is vortex-like. Thus we arrived to the Lagrangian density
\begin{equation}
{\cal L}_{k}=\bar{\Phi}^{k}\gamma ^{\mu }\left( \partial _{\mu }+i\,A_{\mu
}^{k}\right) \Phi ^{k}+m\bar{\Phi}^{k}\Phi ^{k}\,.
\end{equation}
Indeed, the reverse step would be to get rid of the gauge field $A_{\mu }$
by performing a singular gauge transformation 
\begin{equation}
\Phi^{k}(x)\to e^{-i\int_{x_{0}}^{x}dx^{^{\prime }\mu }A_{\mu }^{k}(x^{^{\prime }})}\Phi
^{k}\left( x\right) \;,
\end{equation}
(where $x_{0}$ is an arbitrary fixed point). Since the transformation
is singular, one goes back to a multivalued field. By the same token, we
learn that in order to reproduce the boundary conditions on $\Phi^{k}$,
we should have 
\begin{eqnarray}
\oint_{C_{u_{i}}}dx^{\mu }A_{\mu }^{k}(x) &=&-\frac{2 \pi k}{n} \,,  \label{a}
\\
\oint_{C_{v_{i}}}dx^{\mu }A_{\mu }^{k}(x) &=&\frac{2 \pi k}{n} \,.
\label{b}
\end{eqnarray}
As a side remark, we note that the addition of terms of the form $2\pi
q$, with $q$ an integer, to the right hand side of the above formulae
does not change the total phase factor along the circuits. However,
the winding number of the phase, and thus the boundary conditions the
gauge field imposes, would be different. In fact, the free energy does
depend on these integers and choosing $q\neq 0$ would not select
the vacuum state.

Equations (\ref{a}) and (\ref{b}) hold for any two circuits
$C_{u_{i}}$ and $C_{v_{i}}$  containing $u_{i}$ and $v_{i}$
respectively. Thus:
\begin{equation}
\epsilon ^{\mu \nu }\partial _{\nu}A_{\mu }^{k}(x)=2\pi \frac{k}{n}
\sum_{i=1}^{p}\big[ \delta (x-u_{i})-\delta (x-v_{i})\big] \,,  \label{aa}
\end{equation}
where the presence of a vortex-antivortex pair for each $k$ and each interval is
explicit.  The functional integral is then factorized:
\begin{equation}
Z[n]=\prod_{k=-(n-1)/2}^{(n-1)/2}Z_{k}\,,
\end{equation}
where $Z_{k}$ can be obtained as vacuum expectation values in the free Dirac
theory 
\begin{equation}
Z_{k}=\left\langle e^{i\int A_{\mu }^{k}j_{k}^{\mu }d^{2}x}\right\rangle \,,
\label{fj}
\end{equation}
where $j_{k}^{\mu }$ is the Dirac current, $A_{\mu }^{k}$ satisfies
(\ref {aa}), and we adopted a normalization such that $\left\langle 1
\right\rangle = 1$.

\section{Bosonization and the massless case}\label{sec:massless}
In order to evaluate (\ref{fj}), it is quite convenient to use the
bosonization technique~\cite{bosoandsg}, to express the current
$j_{k}^{\mu }$ as
\begin{equation}
j_{k}^{\mu }\to \frac{1}{\sqrt{\pi }}\epsilon ^{\mu \nu }\partial_{\nu}\phi \,,  \label{corr}
\end{equation}
where $\phi $ is a real scalar field. For a free massless Dirac field, the
theory for the dual field $\phi $ is simply
\begin{equation}
{\cal L}_{\phi }=\frac{1}{2}\partial _{\mu }\phi \partial ^{\mu }\phi \,.
\label{th}
\end{equation}

Therefore we have to evaluate:
\begin{equation}
Z_{k}=\left\langle e^{i\int A_{\mu }^{k}\frac{1}{\sqrt{\pi }}\epsilon ^{\mu
\nu }\partial _{\nu}\phi d^{2}x}\right\rangle =\left\langle e^{-i\sqrt{4\pi }
\frac{k}{n}\sum_{i=1}^{p}\left( \phi (u_{i})-\phi (v_{i})\right)
}\right\rangle \,,
\label{zz}
\end{equation}
where the vacuum expectation values correspond to the
theory (\ref{th}). Since ${\cal L}_\phi$ is quadratic
\begin{equation}
\left\langle e^{-i\int f(x)\phi (x)d^{2}x}\right\rangle =e^{-\frac{1}{2}\int
f(x)G(x-y)f(y)d^{2}xd^{2}y}\,,
\end{equation}
with the correlator 
\begin{equation}
G(x-y)=-\frac{1}{2\pi }\log \left| x-y\right| \,,
\end{equation}
it follows that (\ref{zz}) can be written as
\begin{eqnarray}
\log Z_{k} &=&-\frac{2k^{2}}{n^{2}}\Xi \left( u_{i},v_{j}\right) \,, \\
\Xi \left( u_{i},v_{j}\right) &=&\sum_{i,j}\log
\left| u_{i}-v_{j}\right|-\sum_{i<j}\log \left|
u_{i}-u_{j}\right| -\sum_{i<j}\log \left| v_{i}-v_{j}\right|  -p\log \varepsilon  \,.
\end{eqnarray}
Here $\varepsilon $ is a cutoff introduced to split the coincidence points, $
\left| u_{i}-u_{i}\right| $, $\left| v_{i}-v_{i}\right| \to \varepsilon $. Summing
over $k$ and using (\ref{unos}) and (\ref{entro}) we obtain
\begin{eqnarray}
S_n &=& \frac{1}{1-n}\log (\textrm{tr}\,\rho ^{n}) =\frac{1}{1-n}\sum_{k}\log Z_{k}=\frac{1}{6} \frac{n+1}{n}
 \Xi \left( u_{i},v_{j}\right) \,, \\
S &=&\frac{1}{3}\Xi \left( u_{i},v_{j}\right) \,.  \label{et}
\end{eqnarray}
This agrees exactly with the general formula for the entanglement
entropy for conformal theories obtained in
\cite{Calabrese:2004eu}.  The general case differs from
(\ref{et}) on a global factor of the Virasoro central charge $C$, where
$C=1$ for the Dirac field.

Equation (\ref{et}) has an interesting corollary: recalling the
definition (\ref{mut}) for the mutual information, it follows that, for
non-intersecting sets $A$, $B$ and $C$
\begin{equation}
I(A,B\cup C)=I(A,B)+I(A,C)\,.
\end{equation}
That is, in contrast to the entropy, the mutual information is extensive (in
each of the sets separately) in the conformal case. This curious property
does not hold in the non conformal case or in more dimensions. It can be
written as
\begin{equation}
I(A,B)=\frac{1}{3}\int_{A}dx\int_{B}dy\frac{1}{\left( x-y\right) ^{2}}
\end{equation}
for any two non intersecting $A$ and $B$. This formula for $I(A,B)$ shows 
explicitly its model independent properties, that is, it is cutoff
 independent, positive and monotonically increasing with $A$ and $B$. 

For a single interval we have that
\begin{equation}\label{central}
c_{\alpha} \;=\;\frac{\alpha +1}{6\alpha}\;\;,\;\;\; c\;=\;\frac{1}{3}\,, 
\end{equation} 
which are constants.

\section{The massive case}\label{sec:massive}
We consider here a massive fermion, focusing on the case of a single
interval of length $r$.

We can still use (\ref{corr}) to deal with (\ref{fj}) for the
partition function. Now, however, the bosonization of the massive fermion theory
leads to a sine-Gordon Lagrangian~\cite{bosoandsg}
\begin{equation}
{\cal L}=\frac{1}{2}\left( \partial _{\mu }\phi \partial ^{\mu }\phi +\Lambda
\cos (\sqrt{4\pi }\phi )\right) \,,  \label{la}
\end{equation}
where $\Lambda $ is a mass parameter. Then, as in the previous section we
have 
\begin{equation}
\log \left( \textrm{tr}\,\rho ^{n}(r)\right) =\sum_{k=-(n-1)/2}^{(n-1)/2}\log \left\langle e^{-i\sqrt{
4\pi }\frac{k}{N}\phi (r)}e^{i\sqrt{4\pi }\frac{k}{N}\phi
(0)}\right\rangle \,\,,
\end{equation}
where now the expectation value is evaluated in the sine-Gordon theory
at the free fermion point given by the Lagrangian (\ref{la}). The
correlators of exponential operators were studied in ref. \cite{Bernard:1994re}, where it was shown 
that $\left\langle e^{i\sqrt{4\pi }\alpha
    \phi (r)}e^{i\sqrt{4\pi }\alpha ^{^{\prime }}\phi (0)}\right\rangle $ for $ \alpha $, $\alpha ^{^{\prime
  }}\in [0,1]$ can be parametrized by a function satisfying a nonlinear
second order differential equation of the Painlev\'e type. This is done
by mapping the sum over the form factors into the determinant of a
Fredholm operator. The general relation of this type of determinants
and differential equations is studied in~\cite{widom}. The result
of~\cite{Bernard:1994re,widom} can not be directly applied here, since
we need the correlator for $\alpha ^{^{\prime }}=-\alpha $. This case can be dealt
with through a minor modification in those results (see Appendix A for
more details).

To proceed, one introduces the function 
\begin{equation}
w_{a }(x)=r\frac{d}{dr}\log \left\langle e^{-i\sqrt{4\pi } a  \phi
(r)}e^{i\sqrt{4\pi } a \phi (0)}\right\rangle \,,  \label{dw}
\end{equation}
with 
\begin{equation}
w_{a }(x)=-\int_{x}^{\infty }y\,v_{a }^{2}(y)\,dy \,,  \label{uni}
\end{equation}
where $v_{a }$ satisfies 
\begin{equation}
v_{a }^{^{\prime \prime }}+\frac{1}{x}v_{a }{}^{^{\prime }}=-\frac{
v_{a }}{1-v_{a }^{2}}\left( v_{a }^{^{\prime }}\right)
^{2}+v_{a }-v_{a }{}^{3}+\frac{4 a ^{2}}{x^{2}}\frac{v_{a
}}{1-v_{a }{}^{2}}\,.  \label{q}
\end{equation}
Here we have defined $x\equiv rm$, and the boundary condition for  (\ref{q}) is 
\begin{equation}
v_{a }(x)\sim \frac{2}{\pi }\sin (\pi a )K_{2 a}
(x)\,\,\,\,\,\,\,\textrm{as}\,\,\,\,\,x\to \infty \,,  \label{large}
\end{equation}
where $K_{2 a}(x)$ is the standard modified Bessel function.

Thus, (\ref{uni}), (\ref{q}) and (\ref{large}) give the exact value of $c_n(r)$
 in terms of a finite number of Painlev\'e type functions 
\begin{equation}
c_n(x)=\frac{1}{1-n}\sum_{k=-(n-1)/2}^{(n-1)/2}w_{k/n}(x)\,.
\label{doblew}
\end{equation}

The functions $c_n(x)$ are shown in figure (2) for some values of $n$.
They take the massless case value $\frac{n+1}{6 n}$ at $x=0$, and lie
between $1/6$ for $n=\infty$ and $1/4$ for $n=2$. At large $x=rm$ they
decay exponentially fast. Note that the case $n=1$ corresponds to the
entropic $c$-function, which at the origin takes the value $1/3$ (see
figure 3).

We have also made a direct numerical evaluations of $c_n(x)$ on the
lattice, with a method which is described in Appendix B. Figure 2
shows the results for $n=2$ and $n=3$, which match the exact
theoretical values given by (\ref{doblew}). For higher values of $n$,
the approach to the continuum limit of $c_n(r)$ is slower, and
agreement within few percent requires lattice sets bigger than $1000$
points, already for $n=4$.

\begin{figure}
\centering
\leavevmode
\epsfysize=6cm
\epsfbox{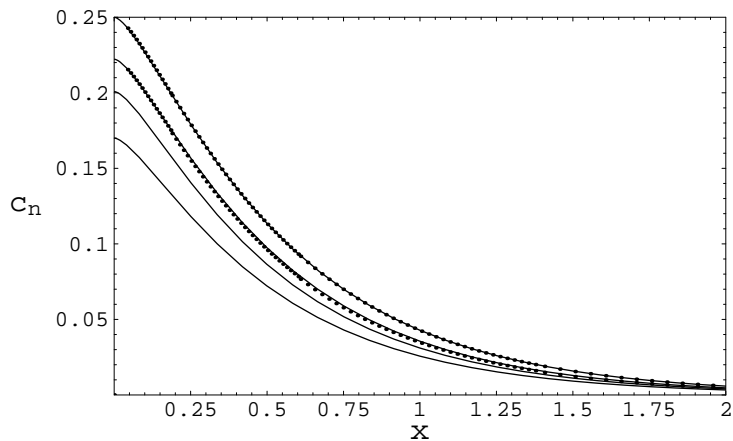}
\bigskip
\caption{Solid lines are plots of  $c_n(x)$, obtained by solving
  differential equations (\ref{uni}-\ref{doblew}) numerically.  The
  values of $n$ are, from top to bottom, $n=2$, $3$, $5$ and $50$.
  These functions take the value $(n+1)/(6n)$ at the origin, which
  cumulates at $1/6=0.166...$ for large $n$. For large $x$, they decay
  exponentially fast.
  Dotted lines correspond to the $c_n(x)$ that results from putting the
  model on a lattice, for $n=2$ and $n=3$. This points are evaluated
  for set sizes ranging from $200$ to $600$ lattice points and inverse
  mass values ranging from $200$ to $3200$ lattice units.  The fact
  that these points, computed for several different lattice mass
  values, tend to lie in a single continuous curve shows the universal
  character of $c_n(x)$.}
\end{figure}

In order to evaluate $c_{n}(r)$ for non integer $n$, or to compute the
entropic $c$ function, we would need an analytical continuation in $n$
of (\ref {doblew}). In what follows we obtain these quantities for
expansions at long and short distances.

\subsection{The long distance expansion}

A naive expansion for large $x$ in (\ref{large}) is not sufficient to 
evaluate the entropy, which requires the $\alpha\to 1$ limit of $c_{\alpha}$.
Indeed, expansions for large $x$ can be obtained from (\ref{large}) by using 
the asymptotic form of the Bessel function for large values of its argument
\begin{equation}
K_{a}(x)\sim e^{-x}\left( \sqrt{\frac{\pi }{2}}\frac{1}{\sqrt{x}}+\sqrt{\frac{
\pi }{2}}\frac{1}{8}(4a^{2}-1)\frac{1}{x^{\frac{3}{2}}}+...\right)\,.
\end{equation}
This gives for the sum of $v^{2}$
\begin{eqnarray}
\sum_{k=-(n-1)/2}^{(n-1)/2} v_{k/n}^{2}(x) &\sim &\left( \frac{2}{\pi }\right)
^{2}\sum_{k=-(n-1)/2}^{(n-1)/2}\sin^2 (\pi \frac{k}{n})K_{\frac{2k}{n}}^2
(r)  \label{sum} \\
& \sim & e^{-2x}\Big[ \frac{n}{\pi }\frac{1}{x}+\big( \frac{1}{12}\frac{n^{2}-4}{
\pi n}+\frac{2\cos (\frac{\pi }{n})}{\pi n\sin^{2}(\frac{\pi }{n})}\big) 
\frac{1}{x^{2}}+...\Big] \,.  \nonumber
\end{eqnarray}
Now we can apply this formula for non integer values of $n=\alpha$.
 This provides a good expansion for fixed $\alpha$ and sufficiently large $x$, from
which the expansion for $c_{\alpha}(x)$ follows by term by term integration 
\begin{equation}
c_{\alpha}(r)\sim \frac{e^{-2x}}{\alpha -1}\Big[ \frac{\alpha}{2\pi }+\big( 
\frac{1}{24}\frac{\alpha^{2}-4}{\pi \alpha}+\frac{\cos (\frac{\pi }{\alpha})}{\pi \alpha\sin
^{2}(\frac{\pi }{\alpha})}\big) \frac{1}{x}+...\Big] \,.  \label{asin}
\end{equation}
However, we see from the way the coefficients in the expansion depend on $\alpha$
that the series does not converge unless $x\gg (\alpha-1)^{-1} $
for $\alpha$ approaching $1$.

Thus, (\ref{asin}) cannot be used to compute the entropy. This can be
repaired using directly the first term in the form factor expansion
for the correlators (see Appendix A) or, equivalently, the integral
representation for the Bessel function
\begin{equation}
K_a(x)=\int_{1}^{\infty }du\,e^{-xu}\; \frac{\Big( u+\sqrt{u^{2}-1}
\Big)^{a}+\Big( u+\sqrt{u^{2}-1}\Big) ^{-a}}{2\sqrt{u^{2}-1}}
\,\label{}
\end{equation}
in expression (\ref{sum}) and summing over $k$ inside the double
integral. We have the following asymptotic expression
\begin{eqnarray*}
c_{\alpha}(x) &\sim &-\frac{2}{\pi ^{2}}
\int_{x}^{\infty }dy\,y\int_{1}^{\infty }du\int_{1}^{\infty }dv\,\frac{
e^{-y(u+v)}}{\sqrt{u^{2}-1}\sqrt{v^{2}-1}}\times  \\
&&\times \left( F_{\alpha}\left( \left( u+\sqrt{u^{2}-1}\right) \left( v+\sqrt{
v^{2}-1}\right) \right) +F_{\alpha}\left( \frac{u+\sqrt{u^{2}-1}}{
v+\sqrt{v^{2}-1}}\right) \right) \,,\label{}
\end{eqnarray*}
where 
\begin{equation}
F_{\alpha}(z)=\frac{1}{4 (1-\alpha)}\left( z-\frac{1}{z}\right) \left( \frac{2\cos \left( 
\frac{\pi }{\alpha}\right) \left( z^{\frac{1}{\alpha}}-z^{-\frac{1}{\alpha}}\right) }{z^{
\frac{2}{\alpha}}+z^{-\frac{2}{\alpha}}-2\cos \left( \frac{2\pi }{\alpha}\right) }+\frac{2}{
z^{\frac{1}{\alpha}}-z^{-\frac{1}{\alpha}}}\right) \,.\label{}
\end{equation}

Surprisingly enough, this function is proportional to a delta function
in the limit $\alpha\to 1$. Indeed, for a fixed $z\neq 1$, $F_{\alpha}(z)\to 0$ as $\alpha\to
1$. However, for $z=1$ this limit is singular. Around $z=1$ and $\alpha=1$
the singularity can be isolated by using polar coordinates in the
$(\alpha-1,z-1)$ plane and expanding in the radial coordinate. The result
is
\begin{equation}
F_{\alpha}(z)=-\frac{\pi ^2}{2}\frac{(\alpha-1)}{ \pi ^{2} (\alpha-1)^2+(z-1)^2  
}  \, \left( 1+{\cal O} \left( (z-1),(\alpha-1)\right)\right)\,.\label{} 
\end{equation}
 Thus, we have 
\begin{equation}
\lim_{\alpha\to 1} F_{\alpha}(z)= -\frac{\pi ^{2}}{2}
\delta (z-1)\,, \label{ffor}
\end{equation}
which yields
\begin{equation}
c(x)\sim -\int_{x}^{\infty }dy\,yK_0(2y)=\frac{1}{2}x\,K_1(2x)\,,\label{cmosca}
\end{equation}
in perfect agreement with the numerical results (see figure (3)).

This term corresponds to the first one in the form factor expansion,
which is due to one soliton-antisoliton pair. In the fermion language
it is the contribution from a single fermion-antifermion pair. It is
interesting to note that (\ref{cmosca}) is $1/4$ at the origin, and
thus a single fermion-antifermion pair contributes more than $75\%$ to
the entanglement entropy function $c(r)$ at all scales.

More terms in the expansion for the entropy can be written in terms of
multiple integrals, by using the form factor series for the
correlators (\ref{hache}). First we expand the logarithm of the
correlator order by order, and then sum over $k$ inside the integrals.
This is easily done since we have a formula analogous to (\ref{ffor})
 \begin{equation}
\lim_{n\to 1} \frac{1}{1-n} \sum^{k=(n-1)/2}_{k=-(n-1)/2} 
\sin^{2 q}\left( \frac{k}{n} \pi\right) z^{\frac{2k}{n}}=-\frac{\pi^{\frac{3}{2}}
\Gamma\left(q-\frac{1}{2}\right)}{2\Gamma (q)}\delta (z-1)\,.
\label{}
\end{equation}
We do not write explicitly more terms of the series here, since the
expression for the multiple integrals, though straightforward to
obtain, do not seem to be particularly illuminating. Besides they are
difficult to evaluate numerically.

\begin{figure}
\centering
\leavevmode
\epsfysize=6cm
\bigskip
\epsfbox{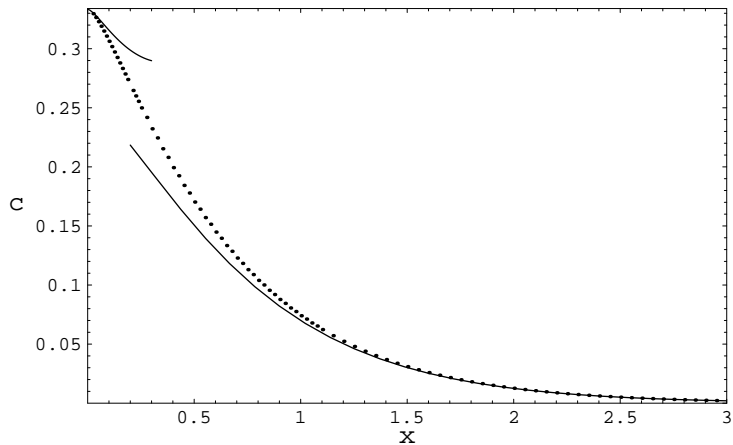}
\caption{The dotted curve is the function $c(r)$ evaluated on a
  lattice, with points obtained for different values of the mass and
  lattice distance (see caption of figure (2)).  The continuity of the
  plot agrees with the universal character of $c(r)$, which is in fact
  a function of $x=mr$. The solid-line curves are the short and long
  distance leading terms we evaluated analytically.}
\end{figure}

\subsection{The short distance expansion}
Close to the conformal limit, the best way to expand $v_{a }(x)$ is by
a direct use of the differential equations. We have the series
solution of (\ref{q}) around the origin
\begin{eqnarray}
v_{a }(x) &=&-2 a \log (x)+b+x^{2}\left( \frac{1}{4}\left( 2 a
-8 a ^{3}+b-8 a ^{2}b-4 a b^{2}-b^{3}\right) +\right.   \nonumber
\\
&&+\frac{1}{2}\left( - a +8 a ^{3}+8 a ^{2}b+3 a
b^{2}\right) \log (x)  \nonumber \\
&&\left. +\left( -4 a ^{3}-3 a ^{2}b\right) \log
^{2}(x)+2 a ^{3}\log ^{3}(x)\right) +\mathcal{O}(x^{4})
\end{eqnarray}
which is of the general form 
\begin{equation}
v_{a }(x)=\sum_{s=0}^{\infty }x^{2s}\sum_{t=0}^{2s+1}f_{s,t}\log
^{t}(x)\,.
\end{equation}
The full expansion requires only the knowledge of the constant term $
f_{0,0}=b$. It might be absorbed into the
logarithm, replacing $\log (x)$ by $\log (xe^{-\frac{b}{2 a }})$ and
setting $b=0$ everywhere in the above expression. Unfortunately we do not
know the general expression for $b$ as a function of $a$. An exception is 
the case $a=1/2$,
where $b=-1-\gamma_E +3\log (2)$, and $\gamma_E$ is the Euler constant. 
 This follows from the results for the Ising
model correlators \cite{Wu:1975mw} (see Appendix A). Here we consider only the
leading term which is independent of $b$. The integration constant for $
w_{a }$ is given by the conformal limit of section 2, $w_{a
}(0)=-2 a ^{2}$. Thus we have 
\begin{eqnarray}
w_{a } &=&-2 a ^{2}+2 a ^{2}x^{2}\log ^{2}(x)+\mathcal{O}
(x^{2}\log (x))\,, \\
c_{\alpha}(x) &=& \frac{\alpha +1}{6 \alpha } \left( 1-x^{2}\log
^{2}(x)\right) +\mathcal{O}(x^{2}\log (x))\,, \\
c(x) &=&\frac{1}{3}-\frac{1}{3}x^{2}\log ^{2}(x)+\mathcal{O}(x^{2}\log
(x))\,.
\end{eqnarray}

\section{Final remarks}\label{sec:summary}
To summarize the results, we have found the exact function $c_{\alpha}(r)$
for a massive fermion field in two dimensions for integer values of
$\alpha=n$. It is given in terms of the solutions of non linear
differential equations. This completely determines the $\alpha$-entropies
for integer $\alpha$, except for a non-universal, ultraviolet divergent,
(additive) constant. We have also found the leading terms of the
entanglement entropy c-function $c(r)$ for short and large distances
showing how it can be expressed as a series of multiple integrals. In
the massless case, we have calculated the entanglement and the
$\alpha$-entropies exactly for an arbitrary set, with results that coincide
with the ones of~\cite{Calabrese:2004eu}.  Surprisingly, the mutual
information turns out to be {\em extensive\/} in the massless case.

There are also other interesting universal quantities which can be
derived from the ones above.  For example, the $\alpha$-entropies increase
logarithmically in the conformal limit,
\begin{equation}
S_\alpha (r)=\frac{\alpha+1}{6\alpha} \log (r) +k_0\,,
\end{equation}
 and saturate for $rm\gg 1$, 
\begin{equation}
S_\alpha (r) \to k_\infty\,.
\end{equation} 
Although the saturation constant $k_\infty$ is cutoff dependent, its dependence on the mass 
is universal and can be expressed in terms of $c_{\alpha}$. Defining an interpolating function  
\begin{equation}
k(r)=s_{\alpha}(r)-c_{\alpha}(r) \log (r)\,,
\end{equation}    
we have $k(0)=k_0$, $k(\infty)=k_\infty$. Using 
$k^{\prime}(r)=-c^{\prime}_{\alpha}(r) \log (r)$ it follows that
\begin{equation}
\Delta k=\int^\infty_0 k^{\prime} (r) dr=-\int^\infty_0 \log(r) c^{\prime}_\alpha(mr) d(mr)
=-\Delta c_\alpha \log (m)+\textrm{const.}\,.
\end{equation} 
This is the result obtained in \cite{Calabrese:2004eu} from the analysis of the properties of the energy 
momentum tensor on a conical space.

There is a very interesting point arising from the equation (\ref{uni}) which
 gives $w_a$ as an integral of an explicitely negative quantity.  
  This implies through equation (\ref{doblew}) that the 
dimensionless functions $c_n(r)$ are always positive and decreasing with $r$ for any 
integer $n\geq 2$.     
 On the other hand, the entanglement entropy c-theorem shows that $c(r)=c_1(r)$ is 
a dimensionless, positive, and decreasing function for {\sl any} two dimensional 
theory \cite{Casini:2004bw}. This seems to suggest that the same is to be expected for the 
 alpha entropy
 c-functions $c_{\alpha}(r)$ for any $\alpha$. This is equivalent to say that
 $S_{\alpha}(r_1)-S_{\alpha}(r_2)$ is decreasing under
 dilatations for $r_1>r_2$, 
\begin{equation}
S_{\alpha}(\lambda r_1)-S_{\alpha}(\lambda r_2) < S_{\alpha}(r_1)-S_{\alpha}(r_2)  \,, 
\label{eki}
\end{equation}
 with $\lambda >1$. 
The corresponding relation for the entanglement entropy ($\alpha =1$) 
 follows from \cite{Casini:2004bw}.

Remarkably, in that case each $c_\alpha$ 
would lead to an alternative form of the c-theorem in two dimensions. Note that 
 the $c_{\alpha}(r)$ are universal quantities which have fixed point values 
proportional to the Virasoro central charge.  

Recently, a different conjecture which also involves the renormalization
 group flow and the reduced density matrices has been presented in a series of papers 
\cite{Vidal:2002rm,Latorre:2003kg,Latorre:2004pk,Orus:2005jq}.
 Those authors proposed several interesting majorization relations for the local density 
matrices and connect them to the renormalization 
group irreversibility. One of the implications of this proposal is 
\begin{equation}
S_\alpha (r_1) > S_\alpha (r_2) \,\,\,\,\,\,\textrm{for} \,\,\,\,r_1>r_2 \,.
\end{equation} 
This means $c_\alpha (r)>0$ in our notation. 
We note that, at least for the entanglement entropy, this relation follows 
from translation invariance only \cite{wehrl}, while (\ref{eki}) with $\alpha=1$ requires 
the full Poincar\'e  group symmetry, which is an essential ingredient 
for the c-theorem~\cite{Zamolodchikov}.

\section{Appendix A: Correlator of exponential operators in the sine-Gordon
model}

The two-point correlation function of (\ref{dw}) can be expanded as a 
sum over form factors as follows \cite{Bernard:1994re}

\begin{eqnarray}
\left\langle :e^{i\sqrt{4\pi } a \phi (r)}::e^{i\sqrt{4\pi } a
^{\prime }\phi (0)}:\right\rangle &=&\sum_{n=0}^{\infty }\frac{1}{\left(
n!\right) ^{2}}\int_{0}^{\infty }du_{1}...du_{2n}\left( \prod_{i=1}^{2n}e^{-
\frac{mr}{2}\left( u_{i}+\frac{1}{u_{i}}\right) }\right) \nonumber \\
&&\times f_{a }(u_{1},...,u_{2n})f_{a ^{\prime
}}(u_{2n},...,u_{1})\,,  \label{hache}
\end{eqnarray}
where $f_{a }(u_{1},...,u_{2n})$ is the form factor
\begin{equation}
f_{a }(u_{1},...,u_{2n})=(-1)^{n(n-1)/2}\left( \frac{\sin \left( \pi
a \right) }{i\pi }\right) ^{n}\left( \prod_{i=1}^{n}\left( \frac{u_{i+n}
}{u_{i}}\right) ^{a }\right) \times \Delta (u_{1},...,u_{2n})\,,
\end{equation}
and
\begin{equation}
\Delta (u_{1},...,u_{2n})=\frac{\prod_{i<j\leq n}\left( u_{i}-u_{j}\right)
\prod_{n+1\leq i<j}\left( u_{i}-u_{j}\right) }{\prod_{r=1}^{n}
\prod_{s=n+1}^{2n}\left( u_{r}+u_{s}\right) }\,.
\end{equation}
This series can be expressed as a Fredholm determinant \cite{Bernard:1994re}
\begin{equation}
\left\langle :e^{i\sqrt{4\pi }a \phi (r)}::e^{i\sqrt{4\pi }a
^{\prime }\phi (0)}:\right\rangle =\det (1-\lambda ^{2}R_{a -a
^{\prime }}R_{a -a ^{\prime }}^{T})\,,
\end{equation}
where $R_{\theta }$ is an integral operator on the half-line $(0,\infty )$
with kernel
\begin{equation}
\left( \frac{u}{v}\right) ^{\theta /2}\frac{e^{\left( \frac{-x}{4}
(u+u^{-1}+v+v^{-1})\right) }}{u+v}\,,
\end{equation}
$R_{\theta }^{T}$ is the transpose of $R_{\theta }$, $x=mr$, and 
\begin{equation}
\lambda =\frac{1}{\pi }\left( \sin (\pi a )\sin (\pi a ^{\prime
})\right) ^{\frac{1}{2}}\,.
\end{equation}
Defining
\begin{equation}
\tau =\log \det (1-\lambda ^{2}R_{\theta }R_{\theta }^{T})
\end{equation}
we have the following differential equations \cite{widom}
\begin{eqnarray}
\frac{d^{2}\tau }{dx^{2}}+\frac{1}{x}\frac{d\tau }{dx} &=&-d^{2}\,,
\label{acc} \\
d^{^{\prime \prime }}+\frac{1}{x}d^{^{\prime }} &=&\frac{d}{1+d^{2}}\left(
d^{^{\prime }}\right) ^{2}+d(1+d^{2})+\frac{\theta ^{2}}{x^{2}}\frac{d}{
1+d^{2}}\,,  \label{ccc}
\end{eqnarray}
with boundary condition 
\begin{equation}
d(x,\lambda )\sim 2\lambda K_{\theta}(x)\;\;\;\;\;\textrm{as}\;\;x\to \infty
\,.
\end{equation}
In the case relevant for this work $a ^{\prime }=-a $, $\theta
=2 a $ and $\,$
\begin{equation}
\lambda =\frac{i}{\pi }\sin (\pi a )\,
\end{equation}
is purely imaginary. This translates into a purely imaginary $d$. Thus,
defining $v_{a }=-i\,d$ equations (\ref{uni}), (\ref{q}) and (\ref{large})
follow.

Equation (\ref{ccc}) has appeared in the literature in different forms. 
It adopts the standard Painlev\'e V form \cite{painleve} by the transformation 
\begin{equation}
d=i\frac{1+h}{1-h}\,,
\end{equation}
which leads to 
\begin{equation}
h^{^{\prime \prime }}+\frac{1}{t}h^{^{\prime }}=\left( \frac{1}{2h}+\frac{1}{
h-1}\right) \left( h^{^{\prime }}\right) ^{2}+\frac{\theta ^{2}}{8}\frac{
(h-1)^{2}}{x^{2}}\left( h-\frac{1}{h}\right) +2\frac{h(h+1)}{(h-1)}\,.
\end{equation}
Also, by the transformation $d=\sinh (f)$ we get the equation 
\begin{equation}
f^{^{\prime \prime }}+\frac{1}{x}f^{^{\prime }}=\frac{1}{2}\sinh (2f)+\frac{
\theta ^{2}}{x^{2}}\frac{\tanh (f)}{\cosh ^{2}(f)}\,,
\end{equation}
which differs however from the equation given in \cite{Bernard:1994re} by a factor $4$ in the
last term. It can be checked that the well-known differential equations for
the Ising model spin and disorder correlators \cite{Wu:1975mw,Fonseca:2003ee}, $\left\langle \sigma
(r)\sigma (0)\right\rangle $ and $\left\langle \mu (r)\mu (0)\right\rangle $, 
can be obtained from the differential equations (\ref{acc}) and (\ref{ccc})
 for the sine-Gordon correlators through the identification \cite{identification} 
\begin{eqnarray}
\left\langle \sigma
(r)\sigma (0)\right\rangle ^2 =\left\langle\sin (\frac{1}{2}\sqrt{4\pi }\phi
(r)) \sin (\frac{1}{2}\sqrt{4\pi }\phi
(0))\right \rangle \,, \\
\left\langle \mu (r)\mu (0)\right\rangle ^2
=\left\langle \cos (\frac{1}{2}\sqrt{4\pi }\phi (r)) \cos (\frac{1}{2}\sqrt{4\pi }\phi (0)) \right \rangle \,.
\end{eqnarray}

\section{Appendix B: Numerical evaluation in the lattice}

The vacuum expectation value $\left\langle O_{A}\right\rangle $ for any
operator $O_{A}$ localized inside a region $A\,$must coincide with tr$(\rho
_{A}O_{A})$, where $\rho _{A}\,$is the local density matrix. This fact was
used in \cite{peschel} to give an expression for $\rho _{A}$ in
terms of correlators for free Boson and Fermion discrete systems. We use this method
here to compute the entanglement entropy for a free Dirac field on a lattice.

Consider a lattice Hamiltonian of the form 
\begin{equation}
{\cal H}=\sum_{i,j}M_{ij}c_{i}^{\dagger }c_{j}\,,  \label{hami}
\end{equation}
where the creation and annihilation operators $c_{i}^{\dagger }$, $c_{j}$
satisfy the anticommutation relations $\{c_{i}^{\dagger },c_{j}\}=\delta
_{ij}$. Let the correlator $C_{ij}=\left\langle c_{i}^{\dagger
}c_{j}\right\rangle $ be taken on any given fixed eigenvector of the
Hamiltonian and call $C_{ij}^{A}$ to the correlator matrix restricted to $A$
. Then the reduced density matrix $\rho _{A}$ has the form 
\begin{equation}
\rho _{A}=\prod \rho _{l}=\prod \frac{e^{-\epsilon _{l}d_{l}^{\dagger }d_{l}}
}{\left( 1+e^{-\epsilon _{l}}\right) },
\end{equation}
where the $d_{l}$ are independent fermion annihilation operators which can be
expressed as linear combination of the $c_{i}$ and $c_{j}^{\dagger }$. The $
\epsilon _{l}$ are related to the eigenvalues $\nu _{l}$ of $C^{A}$ by 
\begin{equation}
e^{-\epsilon _{l}}=\frac{\nu _{l}}{1-\nu _{l}}\,.
\end{equation}
Hence both the entropy and $\log \textrm{tr}\rho ^{n}\,$ can be evaluated as $
S_{A}=\Sigma $ $S_{l}$ and $\log \textrm{tr}\rho ^{n}=\Sigma $ $\log \textrm{tr}
\rho _{l}^{n}$, where 
\begin{eqnarray}
S_{l}\, &=&\log (1+e^{-\epsilon _{l}})+\epsilon _{l}\frac{e^{-\epsilon _{l}}
}{1+e^{-\epsilon _{l}}}=-(1-\nu _{l})\log (1-\nu _{l})-\log (\nu _{l})\nu
_{l}\,, \label{43}\\
\log tr\rho _{l}^{n} &=&\log (1+e^{-n\epsilon _{l}})-n\log (1+e^{-\epsilon
_{l}})=\log ((1-\nu _{l})^{n}+\nu _{l}{}^{n})\,.\label{44}
\end{eqnarray}
We are interested in the vacuum (half filled) state of a Hamiltonian
with symmetric spectrum around the origin. In this case 
\begin{equation}
C=\theta (-M)\,,
\end{equation}
where $\,\theta (x)=(1+$sign$(x))/2$.

The lattice Hamiltonian for a Dirac fermion follows from the discretization
of the Hamiltonian for a Majorana field, to cope with fermion doubling. This is
given by   
\begin{equation}
{\cal H}=\frac{-i}{2}\sum_{n=0}^{N-1}(\Psi _{n}^{1}\Psi _{n+1}^{1}+\Psi
_{n}^{2}\Psi _{n+1}^{2})+i\,m\sum_{n=0}^{N-1}\Psi _{n}^{1}\Psi _{n}^{2}\,,
\end{equation}
with $\Psi^i_n=\Psi^{i \dagger}_{n} $, $\Psi _{N}^{i}=\Psi _{0}^{i}$ and $\{\Psi _{n}^{i},\Psi
_{m}^{j}\}=\delta _{mn}\delta _{ij}$. We have taken $N$ lattice sites and
set the lattice spacing to one. Redefining the fermionic operators as 
\begin{eqnarray}
c_{2k} &=&\frac{1}{2}(\Psi _{2k}^{1}+i\Psi _{2k}^{2})\,, \\
c_{2k+1} &=&\frac{1}{2}(\Psi _{2k+1}^{1}-i\Psi _{2k+1}^{2})\,,
\end{eqnarray}
with $\{c_{n}^{\dagger },c_{m}\}=\delta _{mn}$ we have a Hamiltonian in the
form (\ref{hami}) 
\begin{equation}
{\cal H}=\frac{-i}{2}\sum_{n=0}^{N-1}(c_{n+1}^{\dagger }c_{n}-c_{n}^{\dagger
}c_{n+1}^{\dagger })+m\sum_{n=0}^{N-1}(-1)^{n}c_{n}^{\dagger }c_{n}
\end{equation}
with 
\begin{equation}
M_{ij}=\frac{-i}{2}(\delta (i,j-1)-\delta (i,j+1))+m\,\delta (i,j)(-1)^{n}\,.
\end{equation}
The correlators for the infinite lattice limit $N\to \infty $ are then 
\begin{eqnarray}
\langle c_{i}^{\dagger }c_{j}\rangle  &=&\frac{1}{2}\delta
_{(i-j),0}+(-1)^{i}\int_{0}^{\frac{1}{2}}dx\frac{m\cos (2\pi x(i-j))}{\sqrt{
m^2+\sin (2\pi x)^{2}}}\;\;\;\;\;\;\,\textrm{for} \,\,\,i-j\;\;\textrm{even}\,, \\
\langle c_{i}^{\dagger }c_{j}\rangle  &=&i\int_{0}^{\frac{1}{2}}dx\frac{\sin
(2\pi x)}{\sqrt{m^2+\sin (2\pi x)^{2}}}\sin (2\pi
x(i-j))\;\;\;\;\;\;\,\,\,\,\,\,\,\textrm{for}\,\,\,i-j\;\;\textrm{odd}\,.
\end{eqnarray}
Similar expressions where used in \cite{Latorre:2003kg}.

From (\ref{43}) and (\ref{44}) we can obtain $c_n(r)$ and $c(r)$ by taking numerical
 derivatives. We have used 
\begin{equation}
f^{'}\left( q+ \frac{1}{2} \right)=\frac{1}{4} \left( f(q+2)+f(q+1)-f(q)-f(q-1)\right)\,.\label{}
\end{equation}
This form for the derivative smoothes small oscillations which appear for  
quantities depending on a lattice size $q$ when evaluated between even an odd adjacent values of $q$.

\end{document}